\documentclass[a4,10pt]{article}
\usepackage{a4wide}
\usepackage{subfig}
\usepackage{comment}
\usepackage{graphicx}
\usepackage{amsmath}
\usepackage{amsthm}
\usepackage{amssymb}
\usepackage{amsfonts}
\usepackage[usenames]{color}
\usepackage{appendix}
\usepackage{setspace}
\usepackage{leftidx}
\usepackage{multirow}
\usepackage{cite}
\usepackage{psfrag}
\usepackage[pagebackref=true]{hyperref}
\hypersetup{
    colorlinks=true,       
    linkcolor=blue,          
    citecolor=green,       
    filecolor=magenta,      
    urlcolor=blue          
}

\newtheorem{defi}{Definition}

\newcommand*{\bc}{k_{\textup{B}}}

\newcommand{\one}{($i$) }
\newcommand{\two}{($ii$) }
\newcommand{\three}{($iii$) }
\newcommand{\four}{($iv$) }

\usepackage{array}
\newcolumntype{C}[1]{>{\centering\let\newline\\\arraybackslash\hspace{0pt}}m{#1}}
\title{A thermodynamically consistent model of finite state machines}

\author{
Dominique Chu \\ School of Computing \\ University of Kent \\ CT2 7NF, Canterbury \\ UK
\and 
Richard E. Spinney\\  Centre for Complex Systems \\ The University of Sydney \\ Sydney NSW 2006 \\ Australia}

\usepackage[top=0.75in, bottom=0.75in,left=0.75in,right=0.75in]{geometry}
\begin{document}
\maketitle

\begin{abstract}
Finite state machines (FSMs) are a theoretically and practically important model of computation.  We propose a general, thermodynamically consistent model of FSMs and characterise the resource requirements of these machines.   We  model  FSMs as  time-inhomogeneous Markov chains. The  computation is driven by instantaneous manipulations of the energy levels of the states.  We calculate the entropy production of the machine, its error probability, and the time required  to complete one update step. We find that a sequence of generalised bit-setting operations is sufficient to implement any FSM. 
\end{abstract}

\section{Introduction}

Whenever a computation is done in the real world,  some corresponding physical change to the world   occurs (in Landauer's  words: ``all computation is physical''). This change can be that  work was performed on a system, or entropy produced or some heat generated.   Attempts to  understand the physical consequences of computation   go back to  pioneering work in the 1980s. A key result  that was established then is  the  celebrated  {\em Landauer limit}. It postulates  that writing a bit to a memory device decreases the entropy of the memory by $\bc\ln(2)$ and thus   requires at least an amount of  $\bc T \ln(2)$  work to be done.  Yet, this Landauer limit does not fundamentally bound  the cost of computation. As  was shown  by Bennett   \cite{bennetthermo,timespace}  it is possible to   conceive systems that compute with no lower limits on the thermodynamic  cost.  The caveat is that all  zero-cost computations  proposed so-far either require  infinite time to complete or  return a  wrong results with a high probability. Both characteristics are unacceptable for real-world computations. The cost of computations that complete within finite time and return the correct result with high probability remains an open question. 
\par
Through  relatively recent  theoretical advancements leading to the frameworks of   stochastic thermodynamics \cite{seifertreview} and information thermodynamics \cite{infothermreview}  it has now become  possible to probe rigorously  the physical underpinnings of information processing. Applications of these frameworks to the physics of  information processing  have so far predominantly concentrated on    proto-computations such as bit-rewriting or copying \cite{diana} and  to  biological systems  \cite{wlan,eule,perscopy,government2,uncertain}.  In contrast, there are relatively few  recent attempts to  apply information thermodynamics directly to   computer science. Exceptions are   \cite{wolpertextending,brandes}   who however  consider fundamental limits to the thermodynamic cost of computation in the quasi-static limit. While this  is of high theoretical interest, focussing on the quasi-static  limit, which assumes infinitely slow processes,  misses out on  practically important performance measures of real-world computations, namely  execution  time and accuracy. 
\par
In the present contribution we now wish go beyond quasi-static computations and   probe the thermodynamic cost of computations that complete within a finite time and return the correct result with a probability close to 1.  More specifically, we will not be content with considering proto-computations, or computational processes, but we will insist on staying close to computational theory, so as to understand the resource requirements that are implied by models of computations as they are formulated in computer science. To focus the discussion, we will consider   a particular model of computation, namely {\em finite state machines} (FSM).
\par
In computer science,  models of computation  are mathematically rigorous descriptions of automated mechanisms of symbolic  reasoning. There is not just one such model, but many different ones.  A fundamental result in theoretical  computer science is that  there is a class of  ``Turing complete'' or ``universal''   models that are \one  equivalent to one another and  \two more general than the models to which they are not equivalent. Equivalence   means that anything that can be represented in one model can be translated into the other and {\em vice versa}. Turing complete models  can  represent non-Turing complete models as well, but the reverse is not true.  
\par
A well known example of a complete  model of computation are Turing machines. A Turing machine is a formal construct that consists of a reading head that \one at any one time is in one of a finite number of  possible internal states \two is in contact with a symbol from a tape;  the tape symbols are drawn  from a finite set and the reading head  can  read  the symbol and overwrite it;  finally  \three  it is characterised by   a fixed set of rules. These rules specify the next state of the head depending on the current tape symbol,  what symbol to write to the current tape element (possibly the same as the one encountered), and whether the reading head should move left or right next.    It is  essential to the universality of the Turing machine that the tape is unlimited, i.e.  that the machine has an infinite amount of temporary memory at its disposal. 
\par
In contrast to Turing machines, real-world computers never have unlimited resources.  Turing machines may therefore often not be very good descriptions of such devices. FSMs which only ever consume finite resources  are often a useful  alternative model. They are of particular interest in situations where resources are severely limited, as is the case in, for example, {\em in vivo} computers \cite{fsmindna,fsmindna2,fsmindna3}.  On the downside, they are also non-universal, which means that there are computations that they cannot do. 
\par
Similar to Turing machines, FSMs  have internal states and  can be thought of as having a reading head that can read  symbols from, but not write to,  an external tape. Furthermore, the reading head moves   consistently into one direction on the tape \cite{finitestates} thus receiving a sequence of symbols. Formally, they can be defined as follows:
\begin{defi}[FSM]
A FSM is a tuple $(Q,\Sigma,\delta,I,F)$, where $Q$ is a finite set of internal states, $\Sigma$ represents a finite alphabet of $n$ different  symbols, $\delta$ is a transition function $Q\times\Sigma \rightarrow Q$ and $I$ and $F$ are a set of initial states and  accepting states respectively. 
\end{defi}
The operation of an FSM is as follows:  At time $t=0$ the machine is in a designated start state and in contact with the first symbol on the tape, which will be one of $n$ possible symbols that constitute the tape of length $L$. The transition function  $\delta$ defines  the next state for every state-symbol pair.  State transitions always  coincide with  moving the tape ahead,  such that the FSM is in contact with the next symbol on the tape  when reaching the new state. The machine is updated  until  the end of the tape is reached, at which point the computation halts. If the final state of the machine is a designated ``accepting'' state, then  the input tape is deemed as ``accepted,'' otherwise it is not.
\par
  Computer scientists distinguish between deterministic and non-deterministic FSMs.  A deterministic FSM is one where for each state there is at most one possible transition for each possible tape symbol, i.e. given a tape symbol the next state is determined. Non-deterministic FSMs, on the other hand, may have several possible transitions for each state/symbol pair. It is a well known result that deterministic and non-deterministic FSMs are computationally equivalent, in the sense that the two classes can solve the same problems. Henceforth, we will only consider deterministic FSMs. 
%
%
%
%

\par
One attraction of studying FSMs is that each state transition can be understood as a   {\em unit} of computation, or an {\em elementary cycle}, as we will refer to it. Once we know the thermodynamics of such an elementary unit, we know the thermodynamics of any conceivable FSM because any FSM is just a sequence of those elementary cycles. Furthermore, it will become evident that these elementary cycles can themselves be further broken down into more fundamental computational steps. We will find that only two types of computational steps are required  to implement any FSM, namely  generalisations of bit flips and bit sets that we call $N$-it setters and $N$-it flips. 
\par
The main results of this article are: \one We   show that an elementary cycle of a FSM can be implemented as a sequences $N$-it setters followed by $N$-it flips. This implies that these components, even though they are extremely simple physical mechanisms, are sufficient for a very large class of computational problems.   \two We  present a general scheme to translate any specification of an FSM into a thermodynamically consistent model with a defined thermodynamic cost, and operating accuracy and execution time. \three   We derive a relationship between the cost and error probability per state transition of the FSM  and show that in the limit of high accuracy the resource consumption of an FSM when implemented as a Markovian system,  becomes independent of the size of the tape  alphabet.

\section{Results}

\subsection{Stochastic information processors}
\label{nway}

\begin{figure}
\centering
\subfloat[A $N$-it setter.]{ \includegraphics[width=0.75\textwidth]{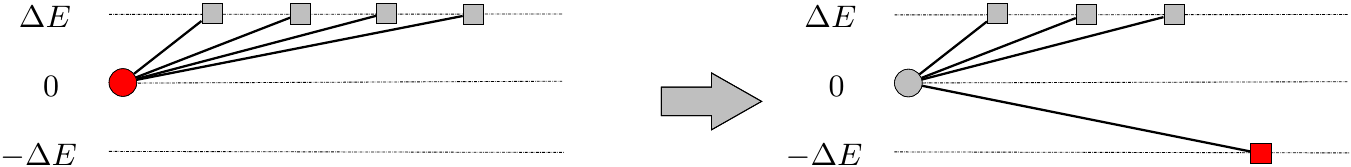} \label{set}}
\\
\subfloat[An $N$-it flip.]{\includegraphics[width=0.75\textwidth]{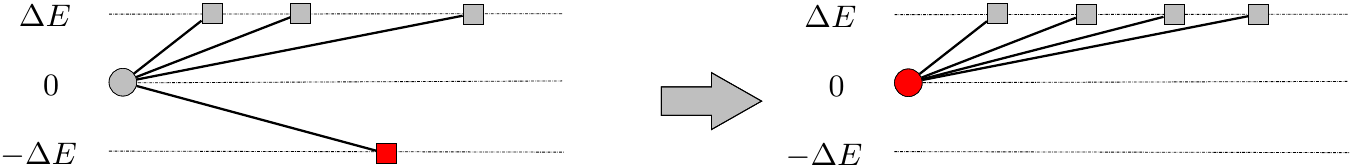} \label{flip}}
\caption{Schematic representation of a $N$-it setter \protect\subref{set} and $N$-it flip \protect\subref{flip} for $N=4$. The circle represents the state $L$ and the states $D_i$ are indicated by the squares. The preferential state is indicated in red.  Transitions are only allowed between states connected by lines.  For the $N$-it setter the initial  state  $E=0$. The target state to be set is then lowered to energy $E=-\Delta E$, which sets the state. The $N$-it flip is the reverse procedure, where the system starts out in a state with energy $E=-\Delta E$, which is then lifted to $E=\Delta E$, at which point the state at $E=0$ is taken.}
 \label{nitsetter}
\end{figure}

Before describing a physically plausible model of FSMs, we will introduce the two atomic processes that we will act as  fundamental building blocks from which arbitrary FSMs can be built and describe their relevant thermodynamics. Specifically these are generalisations of simple bit-set and bit-flip operation, but with multiple states and only three energy levels. We call these $N$-it setters and $N$-it flips.
\par
Firstly, however, for clarity, we consider a simple bit-flip and bit-set as often conceived of, modelled by a two state system, $\{1,2\}$. First we consider the bit flip where each state is initially associated with an energy level $\{-\Delta E,0\}$. In this model the system is also in contact with an external heat-bath that is kept at some constant  temperature $T$. On sufficiently long timescales the system will be found in state $1$ with probability $\pi^i_1 = \exp(\Delta E/\bc T)/(1 +\exp(\Delta E/\bc T))$, $i$ signifying initial. The energy levels correspond to transition rates between the two states, namely $k_{1\to 2}$ and $k_{2\to 1}$, which we initially set as $k_{1\to 2}=k^-$ and $k_{2\to 1}=k_+$, $k_+>k_-$, through a postulate that the system obeys local detailed balance \cite{localdetailedbalance}, i.e.   $k_+/k_-=\exp(\Delta E/kT)$. To enact the bit flip the energy level of state one is raised to $E_1=+\Delta E$, upon which the transition rates exchange values such that $k_{2\to 1}=k^-$ and $k_{1\to 2}=k_+$ and after a suitably long time, the system will be found in state 2 with probability $\pi^f_2 = \pi_1^i$, $f$ signifying final. To enact the bit flip thus, on average, incurs work 
\begin{align}
\langle\Delta W\rangle=2\Delta E\pi_1^i=\frac{2k_+}{k_++k_-}\ln\left(\frac{k_+}{k_-}\right)\geq 0
\end{align}
and total entropy production
\begin{align}
\Delta S_{\rm tot}=\langle\Delta W\rangle-\Delta F=\langle\Delta W\rangle-\Delta E=\frac{k_+-k_-}{k_++k_-}\ln\left(\frac{k_+}{k_-}\right)\geq 0
\end{align}
where we have set  $\bc,T=1$ here and throughout. This is easily computed by considering the difference in the free energy $F=-\ln Z$ in the initial and final configurations where $Z=\sum_{i\in{\rm states}}e^{-E(i)}$ is the partition sum.
\par
In contrast a bit set operation, modelled here as a relaxation, transforms a distribution governed by the energy landscape $\{0,0\}$ into one governed by $\{-\Delta E,0\}$. As such the probabilities evolve from $\pi_1^i=1/2$ to $\pi^f_1=\exp(\Delta E)/(1+\exp(\Delta E))$. Consequently the work associated with the bit-set is
\begin{align}
\langle\Delta W\rangle=-\Delta E\pi_1^i=-\frac{1}{2}\ln\left(\frac{k_+}{k_-}\right)\leq 0
\end{align}
and the mean total entropy production is
\begin{align}
\Delta S_{\rm tot}=\langle\Delta W\rangle-\Delta F=\ln\left(\frac{k_++k_-}{2\sqrt{k_+k_-}}\right)\geq 0.
\end{align}
\subsubsection{The $N$-it setter}
\label{bset}
Here we consider the first atomic process  and generalisation of the above bit set, which we call  the $N$-it setter. As with the bit set and bit flip  operations it involves altering the energy landscape in order to drive the system towards a preferential state, exchanging energy as work. However, this operation is slightly more complicated, involving $N+1$ states, where $N$ is a free parameter of the $N$-it setter that we discuss later in the paper. We denote these states $\{L,D_1,\ldots,D_N\}$ for reasons that will become apparent later. As before we assign each state an initial energy level, specifically, $\{0,\Delta E,\ldots,\Delta E\}$. The exchange of work and relaxation we consider then corresponds to the final energy levels, given by $\{0,\Delta E,\ldots,-\Delta E,\ldots,\Delta E\}$ corresponding to states $\{L,D_1,\ldots,D_m,\ldots,D_N\}$ such that state  $D_m$, $1\leq m\leq N$, is preferentially selected or `set' in the operation. Consequently, we are performing a setting of the system into state $D_m$ from state $L$ in the presence of $N$ `distractor' states which reside at a higher energy. As before, energy level differences of $\Delta E$ are effected through transition rates e.g. $k_{L\to D_1}=k_-$, $k_{D_1\to L}=k_+$. Whilst it is of no consequence here, later we require at no time are transitions between different states $D_i$ allowed.
\par
Analogously to the bit flip we may compute the initial probability to be in state $D_m$, given here by $\pi^{i,{\rm set}}_{D_m} = \exp(-\Delta E)/(1 +N\exp(-\Delta E))$. Following the exchange of energy levels and a suitable period of equilibration the final probabilities change with the probability of being in the preferred state $D_m$ now given by 
\begin{align}
\pi^{f,{\rm set}}_{D_m} &= \frac{\exp(\Delta E)}{1+\exp(\Delta E)+(N-1)\exp(-\Delta E))}\nonumber\\
&=\frac{k_+^2}{k_+^2+k_-k_++k_-^2(N-1)}.
\end{align}
 Consequently, we may thus compute the mean work incurred in such an operation as
\begin{align}
\langle\Delta W\rangle=-2\Delta E\pi_{D_m}^{i,{\rm flip}}=-\frac{2k_-}{k_++Nk_-}\ln\left(\frac{k_+}{k_-}\right)\leq 0
\end{align}
and the entropy production 
\begin{align}
\Delta S_{\rm tot}&=\langle\Delta W\rangle-\Delta F\nonumber\\
&=-\frac{2k_-}{k_++Nk_-}\ln\left(\frac{k_+}{k_-}\right)+\log\left(1+\frac{k_+}{k_-}+\frac{k_-(N-1)}{k_+}\right)-\log\left(1+\frac{Nk_-}{k_+}\right)\geq 0
\label{ent1}
\end{align}

\subsubsection{The $N$-it flip}
\label{bflip}
Here we consider the second atomic operation and generalisation of the above bit flip. Again this operation performs work by altering the energy landscape in order to drive the system towards a preferential state, here driving it out of an already preferential state. As with the $N$-it setter this operates on states $\{L,D_1,\ldots,D_N\}$. However, in this instance we `flip' from the single state $D_m$ to a single target state, $L$, in the presence of $N-1$ distractor states at a higher energy level. As before, to achieve this, we assign each state an initial energy level, specifically, $\{0,\Delta E,\ldots,-\Delta E,\ldots,\Delta E\}$ corresponding to states $\{L,D_1,\ldots,D_m,\ldots,D_N\}$ such that, again, $D_m$ is the set state with energy level $-\Delta E$. The injection of work and relaxation we consider then corresponds to the final energy levels, given by $\{0,\Delta E,\ldots,\Delta E\}$. Again transitions between $D_i$ states are disallowed (see fig. \ref{nitsetter}).
\par
Analogously to the bit flip we may compute the initial probability to be in the mechanically raised state $D_m$ given here by $\pi^{i,{\rm flip}}_{D_m} = \exp(\Delta E)/(\exp(\Delta E)+1+(N-1)\exp(-\Delta E))$. Following the exchange of energy levels and a suitable period of equilibration the final probability of being in the specific set state $L$ is 
\begin{align}
\pi^{f,{\rm flip}}_{L}&=\frac{1}{(1 +N\exp(-\Delta E))}\nonumber\\
&=\frac{k_+}{k_++Nk_-}. 
\end{align}
Consequently, we may thus compute the mean work incurred in such an operation as
\begin{align}
\langle\Delta W\rangle=2\Delta E\pi^{i,{\rm flip}}_{D_m}=\frac{2k_+^2}{k_+^2+k_+k_-+k_-^2(N-1)}\ln\left(\frac{k_+}{k_-}\right)\geq 0
\end{align}
and the entropy production is given by
\begin{align}
\Delta S_{\rm tot}&=\langle\Delta W\rangle-\Delta F\nonumber\\
&=\frac{2k_+^2}{k_+^2+k_+k_-+k_-^2(N-1)}\ln\left(\frac{k_+}{k_-}\right)+\log\left(1+\frac{Nk_-}{k_+}\right)-\log\left(1+\frac{k_+}{k_m}+\frac{(n-1)k_-}{k_+}\right)\geq 0.
\label{ent2}
\end{align}

\subsection{Timescales of the atomic operations}

In order to understand any trade off between time, energy requirements and accuracy of the FSM we need to understand the characteristic relaxation times to equilibrium following the initial exchange of work in the atomic operations. This is found by solving the relevant master equations associated with the $N+1$ state systems involved in each with transition rates matched to the subsequent energy levels, but with an initial condition matched to the initial energy levels.
Doing so reveals a characteristic time scale for the bit-flip
\begin{align}
\tau^{-1}_{\rm flip}(N)=k_++Nk_-
\end{align}
and a characteristic time for the bit-set of
\begin{equation}
\label{nittime}
\tau^{-1}_{\rm set}(N) = k_++\frac{k_-(N+1)}{2}-\frac{1}{2}\sqrt{k_-(k_-(N-1)^2+4k_+N)}
\end{equation}
In the high accuracy limit,  $k_+\gg  k_-N$,  both $\tau_\textrm{flip}(N)$ and $\tau_\textrm{set}(N)$ scale like  $1/k_+$, independently of $N$. This means that the $N$-it setter and flip procedures operate fast when they operate accurately. 
\par 
The characteristic time for $N$-it setters to reach their equilibrium will  determine the operating time of the FSM. In a strict sense,  the system only reaches its  equilibrium  after an  infinite time.  Yet, equilibrium is approached exponentially fast, thus,  in practice it will be sufficient to wait for a  few multiples of $\tau_{\rm flip}$ or $\tau_{\rm set}$ in order to be  so close to the true equilibrium that  for all practical purposes the probabilities of the system are  indistinguishable from an equilibrium system. If we decide to let the system relax for given multiples of $\tau_{\rm flip}$ or $\tau_{\rm set}$, say 10, before the system is measured (or further manipulated), then the Markov $N$-it setter is operated in a defined finite time.

\subsection{A physical model of FSMs }

We will now turn to constructing a thermodynamically consistent  model of  FSMs.  We  will show that arbitrary FSMs can be constructed  using  a series of $N$-it setters  and $N$-it flips. FSMs are time-inhomogeneous Markov chains, in that the transition rates between two states depend on the current tape-symbol. Therefore, any physical implementation of an FSM also requires a protocol to switch the transition rates between states depending on the tape-symbol. Before  describing the model in section \ref{model}, we will motivate our modelling choices in a step by step manner.    
\par  
We start by  re-interpreting  the above defined  logical model of the  FSM as an (continuous) time-inhomogeneous Markov chain  $\mathcal M$. We do this by associating a single physical state, which we call a \emph{label state}, with each internal state in $Q$ of the FSM. We  refer to these states with symbol $L_i$ where $1\leq i\leq|Q|$. We will then find that this naive translation of the conceptual FSM is not sufficient to implement  the required mechanisms as a credible physical system. It will turn out that one way to solve this problem is to introduce additional  ``helper''  states to the FSM.
\par
We first   translate naively the FSM into a Markov chain model $\mathcal M$. This can be done as follows:
\begin{enumerate}
\item
The states and transitions of $\mathcal M$, with states identified as the set of label states $L_i$, are the same as the  states and transitions of  the FSM.  
\item
Every transition $t_{ij}^a$ from some state $L_i$ to some state $L_j$  is (conceptually) associated with a symbol $a$ of the input alphabet. 
\item
Each time an internal state transition happens, the external tape moves forward and the machine is in contact with the next tape element which contains a (possibly different)  symbol $a$.
\item
When the current tape symbol is $a$  and  if the transition from state $L_i$ to state $L_j$ is associated with $a$ then  the transition rate is $k_{{ij}}=1$ and   $k_{{ji}}=0$; otherwise  $k_{{ij}}=0$.
\end{enumerate} 
\par 
As a next step, we now consider how this abstract Markov chain could be implemented as a plausible physical system. To this end we  re-interpret the chain  as a microscopic system in contact with some heat-bath. Each state $L_i$ of the chain is represented by an energy level $E_i$.  Again postulating local detailed balance we can now derive the transition rates between any  state $i$ and $j$ as  $k_{{ij}}/k_{{ji}}= \exp((E_i-E_j)/ T)$. Note, that this entails that  if the forward transition $L_i\to L_j$ happens, then the reverse transition $j\to i$  may also  take place, albeit with a possibly very small rate. In the original FSM model reverse transitions were  not allowed. The  micro-reversibility is an immediate consequence of thermodynamic consistency.

\subsubsection{Dynamic states are necessary to consistently encode transition rates by energy differences}
Before we arrive at a complete model of the FSM further states, beyond the label states associated with the internal states of the FSM, need to be introduced (see in this context \cite{minimalhidden}). This arises because transitions between states need to be forced by energy differences between the states, i.e. desired  next states should have a low energy and all other states a high energy, but as we will show, this is not possible.
\par
\begin{figure}
\includegraphics[width=0.45\textwidth]{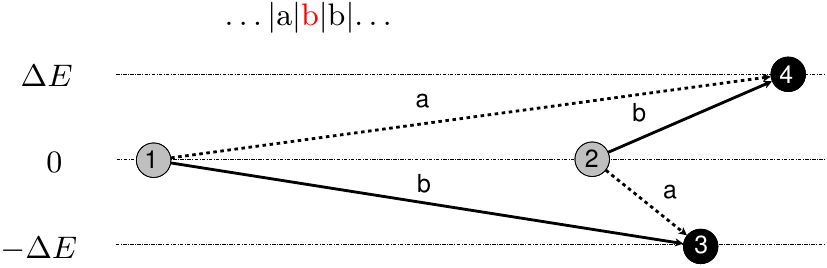}
\centering
\caption{If transitions were from transition states to label states directly, then inconsistencies may appear. Black circles indicate label states, grey circles transition states. The dashed horizontal lines indicate the  three energy levels of the states, which are $\{0, \pm\Delta E\}$. The tape is indicated with the current position in red. The arrows indicate transitions, and the letter above the arrow indicates the symbol with which the transition is associated. Here, the transition associated with symbol {\tt b} from state 1 should be to state 3, whereas the transition from state 2 should be to state 4. This cannot be satisfied simultaneously. Dynamic states avoid this problem.}
 \label{transStates}
\end{figure}

  To see this consider the specific transition from state $L_i$ to state $L_j$. Assume that for symbol $a$ this transition should happen at a low rate $k_-$, i.e. $a$ is associated with a  transition  from $L_i$ to a different state $L_l$. Consequently, the energy of state $L_j$ should be high. At the same time, it  may  be the case that the tape symbol $a$ is associated with  the transition  from a different  state $L_k\neq L_i$  to $L_j$, i.e. this transition should happen with a rate     $k_+$. Accordingly, the energy of $L_j$ should be low.  In general, it is therefore not possible to achieve consistent transition rates by adjusting the energy levels of label states; see fig. \ref{transStates} for a graphical explanation. 
\par
 This can be overcome by  introducing  a new state  for each individual transition between each logical state of the FSM, including, importantly, self transitions. We call these  {\em dynamic states}.  Since every possible tape symbol induces some transition in the FSM (albeit possibly a self transition) this entails adding another $n$ states for each label state, where $n$ is the number of tape-symbols. We denote these states $D^{i\to j}_\alpha=D^{i\to i(\alpha)}_\alpha$ to be interpreted as the dynamic state associated with the transition away from state $L_i$, towards state $L_j$, when symbol $\alpha$ is on the tape. Note that for every $i$ and $\alpha$ combination there is exactly one value $j$, also denoted $j=i(\alpha)$, which may be equal to $i$ in the case of a self transition in the FSM.
 \par
\begin{figure}
\centering\includegraphics[width=0.95\textwidth]{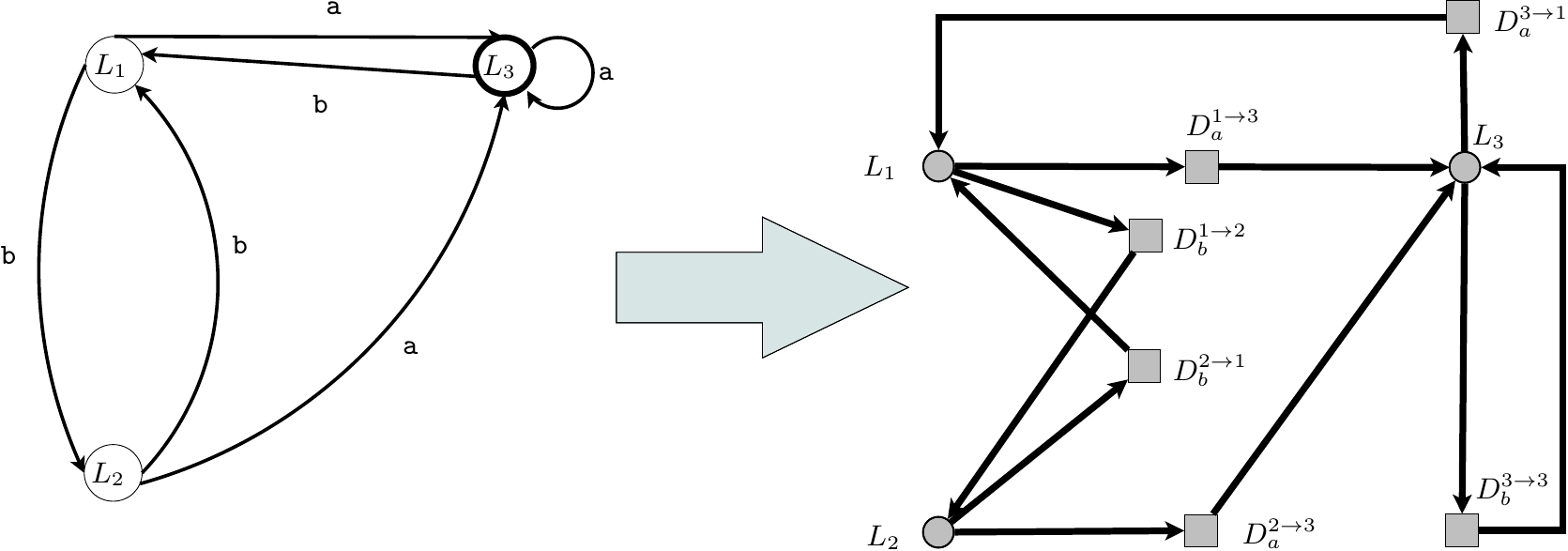}
\centering
\caption{Example of a FSM. ({\bf left}) The logical transition diagram
of a FSM with 3 states. The final/accepting state is state 3. This FSM
recognises tapes that end with the symbol {\tt a}. ({\bf right}) The
state diagram of the physical implementation of the same FSM. Label
states are states 1,2,3.  The dynamic states are indicated by the state
to which they belong and the symbol to which they correspond, i.e. 1a
is the dynamic state for state 1 corresponding to symbol a. The arrows
indicate the intended direction of computation (when the corresponding
symbol is read from the tape); reverse transitions are possible too,
but not indicated. See main text for details. }
\label{transitionfull}
\end{figure}

 Consequently we thus have distinct states for each of the internal states of the FSM (label states), but also for each of the possible `directions' that the system can transition in from each internal state (dynamic states). If there are $M=|Q|$ internal states, then there are $M(n+1)$ total states and $M$ dynamic states corresponding to each possible symbol on the tape. We use notation $D_\alpha$ to denote all dynamic states associated with tape symbol $j$, i.e. $D_\alpha=\{D_\alpha^{1\to 1(\alpha)},\ldots,D_\alpha^{M\to M(\alpha)}\}$ and  $D^i$ to denote all dynamic states associated with label state $L_i$, i.e. $D^i=\{D_a^{i\to i(a)},D_b^{i\to i(b)},\ldots\}$. Any individual label state, $L_i$, will have connections to $n$ outgoing dynamic states, $D^i$, designed to take the system from that label state to another state, and up to $n|Q|$ incoming dynamic states from other label states from which the system is intended to transition. 
%
We summarise the procedure to  transform an FSM with $n$ tape symbols into a physically plausible Markov model:
\begin{enumerate}
\item
Identify every internal state of the FSM with a physical label state, $L_i$.
\item
For each label state, $L_i$, introduce another $n$ states --- the dynamic states, $D^i$. Each  is associated with a different  symbol of the tape alphabet and can connect to exactly one other label state.
\item
Introduce possible  transitions  between  different label states via a dynamic state.  The transitions should be as   specified by the FSM model: If the FSM specifies that there is a transition from state $L_i$ to state $L_j$ for a tape symbol $a$, then the model should contain  reversible  transitions from  the label state $L_i$ to a dynamic state $D^{i\to i(a)}_a=D^{i\to j}_a$ and reversible transition from that dynamic state and the label state $L_j$. 
\end{enumerate} 
Figure \ref{transitionfull} illustrates the procedure with a specific example. 

\subsection{Detailed description of the model}
 \label{model}
In  order to implement any FSM it is sufficient to postulate  three energy levels, corresponding to $E = \{\pm\Delta E,0\}$. We choose that  the  transition and label states always remain at $E=0$. In contrast the dynamic states are at $E=\pm\Delta E$. An {\em elementary cycle} is  the procedure that needs to be executed in order to implement  a single state transition of a FSM. This elementary cycle is generic in that it can be used in any network topology and for  any number of tape symbols; fig \ref{switching} illustrates an elementary cycle for a tape consisting of 2 symbols. It is important to stress that an elementary cycle is not a traditional thermodynamic cycle, such as for example a Carnot cycle, that brings the system back into its initial state. Instead, it is a sequence of manipulations that achieve an elementary logical operation. In general an elementary cycle will not bring the FSM back to the same state.

\subsubsection{Outline of the elementary cycle}

The elementary cycle consists of the 2 atomic computational steps outlined earlier interleaved alongside minimal mitigation to ensure evolution in the correct direction. The purpose of the generalised flips is to extract the system from some dynamic state, $D_\alpha^{i\to j}$ and into the next label state $L_{j}$ as determined by the behaviour of the FSM. The purpose of the generalised sets is then to take the system from a label state $L_j$ into the appropriate dynamic state, $D^{j\to j(l)}_\alpha$, given symbol $\alpha$ is on the tape. 
\par
The mitigation against backwards evolution takes the form of energy barriers between the states. We assume that the heights of  these barriers are  effectively infinite,  such that over relevant time scales there will be no transitions over these barriers. These barriers can be placed between the label and dynamic states in two different configurations. In the first configuration, the \emph{outgoing configuration}, they lie between label states and all outgoing dynamic states associated with them,  i.e. between all states $L_i$ and corresponding dynamic states $D^i$. In the second configuration, the \emph{incoming configuration}, they lie between between the label states and incoming dynamic states that are directed towards them, i.e. between all state $L_j$ and corresponding dynamic states $D^{l\to j}_\alpha$ for all corresponding internal states $l$ and tape symbols $\alpha$. 
\par
Use of such energy barriers amounts to a coarse graining of certain states (or operations) required of the system and thus inevitably results in a model with lower energetic costs. In this sense, our results provide a minimum bound on energy requirements and performance. We do this to simplify the discourse, but utilise minimal use of such barriers in order to retain pertinent features that effect energy input and accuracy -- specifically dependence on alphabet and machine size. We achieve this by insisting that we are only able to arrange them in the outgoing and incoming configurations such that the relevant dependence on the size of the machine and alphabet, reflected in the relevant multiplicities of the $N$-it set and flip operations, are in turn reflected in the cost and precision calculations.
\par
\begin{figure}
\includegraphics[width=\textwidth]{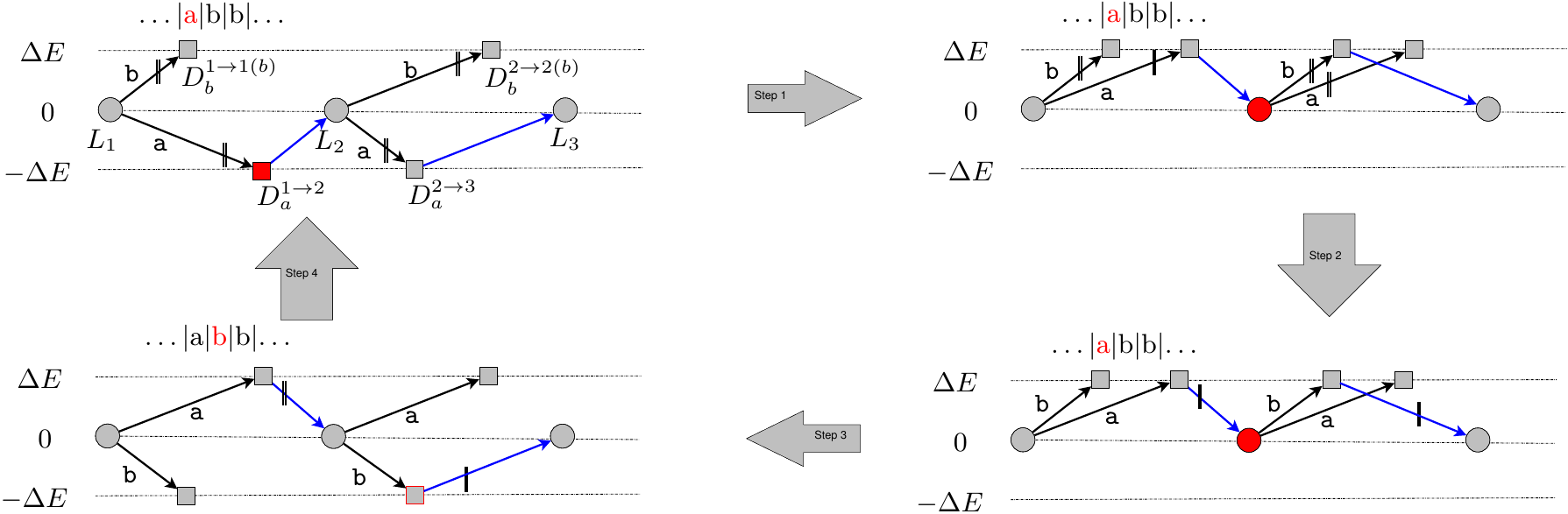}
\centering
\caption{Schematic outline of an elementary cycle; see main text for an explanation for each step. The dashed horizontal lines indicate the  three energy levels of the states, which are $\{0, \pm\Delta E\}$.  The gray circle indicates a label state and the grey square indicates a transition states. For each label state there are $n$ transitions state. For clarity we color the connection from transition states to label states blue. The input tape is shown  with the current symbol being marked in red.  Similarly, the current intended state of the FSM is marked in red. The names of the states are indicated below the figure.  The procedure is followed along the large grey arrows. We start in a state where  the system is in a transition state. During the first step we lift all transition states to the highest energy state.  During the second step we re-distribute the energy barriers from the blue arrows to the black arrows. In the third step  the tape is moved on and the transition states corresponding to the new tape element are lowered. The fourth step takes the system   back  to its original configuration.} \label{switching}
\end{figure}

The machine operates over three fixed energy levels. Continuous operation is therefore only possible if, through input of work,  the energy levels of states are lifted  to drive the  computations. We will now describe a generic protocol so as to realise an arbitrary FSM. Note that  the execution of this protocol does not depend on  knowledge of the state of the machine at a particular time.
\paragraph*{Step by step procedure}
We arbitrarily define the beginning of the   cycle as a machine state where  \one the   label states $L_i$ have an associated energy $E=0$, \two   the machine is in some dynamic  state $D^{i\to i(j)}_\alpha$ corresponding to label state $L_i$ and symbol on the tape $\alpha$. All dynamic states $D_\alpha$ have associated energy $E=-\Delta E$, whilst all other dynamic states, $D_{\alpha'\neq \alpha}$, have associated energy $E=+\Delta E$; see fig. \ref{switching} and \three the energy barriers are in the outgoing configuration.  The machine may be in a different state initially, which may result in an error of the computation; see section \ref{errordiscuss}.
\par
  {\bf Step 1}  Lift all dynamic states  to energy $E=+\Delta E$. Wait for a chosen multiple of $\tau_\textrm{flip}$ before proceeding to the next step.
  This implements a $N$-it flip from section \ref{bflip} relaxing the system into label state $L_j$ to which the currently occupied dynamic state, $D^{\cdot\to j}_{\cdot}$ is directed. Because of the outgoing energy barriers, up to $n|Q|$ possible incoming dynamic states lead into the same label state. Consequently, the bit-flip is performed with $N\leq n|Q|$.
  \par
   {\bf Step 2}    Existing energy barriers are removed and instead new energy barriers are set in the incoming configuration.
   \par
    {\bf Step 3} The tape is advanced, at some cost, and the dynamic state energy levels are set in accordance with the symbol on the read head. Wait for a chosen multiple of $\tau_{\rm set}$ before proceeding to the next step.
    \par
   This implements the $N$-it setter from section \ref{bset} causing the system to enter the dynamic state $D^{i\to i(k)}_\alpha$ associated with the current label state $L_i$ and the symbol $\alpha$ on the tape. Specifically, if symbol $a$ is now on the tape all dynamic states $D_a$ are lowered to energy level $E=-\Delta E$ whilst the rest remain at energy level $E=+\Delta E$.  Because of the incoming energy barriers and the structure of FSMs there are exactly $n$ outgoing dynamic states attached to the current label state ensuring that the bit-set is performed with $N=n$.  
    \par
    {\bf Step 4} Existing energy barriers are removed and instead new energy barriers are set in the outgoing configuration. This takes the machine back to the beginning of a new elementary cycle. 

\subsection{Energetics  of the FSM}

The computation is driven primarily by lifting  the energy levels of the machine during steps 1, with the potential for (small) work extraction in step 3. On average this  requires positive work expenditure.  During step 1 the dynamic state is lifted from $E= -\Delta E$ to $E= \Delta E$. Since step 1 is associated with a $N$-it flip, with $N=N_Q\leq n|Q|$, we associate it with work expenditure
\begin{equation}
\langle\Delta W_1\rangle = \frac{2k_+^2}{k_+^2+k_+k_-+(N_Q-1)k_-^2}\ln\left(\frac{k_+}{k_-}\right).
\end{equation}
Similarly, the work that can be extracted during step three is associated with a $n$-it setter with $N=n$  equals
\begin{equation}
\langle\Delta W_2 \rangle=   -\frac{2k_-}{k_++nk_-}\ln\left(\frac{k_+}{k_-}\right).
\end{equation}
The sum is therefore the total expenditure of work expected under this model
\begin{align}
\langle \Delta W\rangle &= \langle\Delta W_1\rangle + \langle\Delta W_2\rangle\nonumber\\
&=\left(\frac{2k_+^2}{k_+^2+k_+k_-+(N_Q-1)k_-^2}-\frac{2k_-}{k_++nk_-}\right)\ln\left(\frac{k_+}{k_-}\right)\geq 0\nonumber\\
&=2\ln\left(\frac{k_+}{k_-}\right)+\mathcal{O}\left(\frac{k_-}{k_+}\ln\left(\frac{k_-}{k_+}\right)\right).
\label{avgwork}
\end{align}
Specifically, we note in this high accuracy, $k_+\gg k_-$, limit the result is independent of the size of the alphabet $n$ and the size of the machine expressed through $N_Q$.
\par
Note also, that if the energy barriers always enclose the particle (also valid only in the high accuracy limit, $k_+\gg k_-$), and the system returns to the same label state $L_i$ as it was previously, then the system has performed a bonfide thermodynamic cycle captured entirely accurately by these two operations  such that the mean entropy production is the sum of the entropy productions of the atomic operations; this is equal to the work since the free energies cancel, as they must, due to the cyclical operation.
\par
In addition to the energy expenditure during elementary cycles, there is an addition expenditure required to reset the machine to the start state at the beginning of the computation. We note that the  thermodynamic cost of this procedure will depend on the number of states of the FSM, but is independent of the number of computational steps that will be done. In the limit of long tapes $L\gg 1$ it will be  insignificant.

\subsection{Error probabilities}
\label{errordiscuss}
We will now consider the error probabilities implied by the model. For this purpose, we will limit the discussion to errors that arise from the micro-reversibility of the model. Specifically, we will ignore the possibility that the machine may erroneously transition between unconnected states. We also assume that energy barriers cannot be overcome. 
\par
With this caveat in mind,  it is clear  that the   probability of  errors during the computation depends on the energy differences $\Delta E$ between states. The larger the difference, the smaller the error.  The steps where errors occur are, however, not those where the energy levels are changed, but instead those where the barriers are re-distributed.  The re-distribution may  fix the system in incorrect states and thus affect the reliability of the computation.  When analysing the error probability of the machine, we therefore only need to consider   steps 2 and 4. 
\par
 During step 2  the machine could be in  one of the $N_Q\leq n|Q|$ dynamic states (at energy level $\Delta E$) rather than in the intended  label state (at energy level $0$). This will then result in an error. The probability of being in one of these dynamic states, recalling the local detailed balance relation $k_+/k_-=\exp(\Delta E)$, is given by 
\begin{equation}
p_\textrm{err1} = {N_Qk_-\over N_Q k_- + k_+}
\end{equation}
Assuming that no further errors happen, the outcome of this error will be that the machine misses a transition, while the tape moves ahead --- tape and FSM de-synchronise. Under some circumstances this  remains inconsequential, but typically it will mean that the computational path has been altered and the result of the computation is unreliable. 
\par 
Energy barriers are also  re-arranged  during step 4 when another misstep can happen.  The intended state  of the machine before step 4 is a low energy dynamic state (energy level $-\Delta E$). In reality, the machine may be in the relevant label state $L_i$ (energy level $0$) or in  one of the remaining $(n-1)$ high energy dynamic states $D^i$ (energy level $+\Delta E$). The  probability $p_\textrm{err2}$ is the probability of being in any state other than the low energy dynamic state, i.e.
\begin{equation}
p_\textrm{err2}= 1- {\frac {k_+^{2}}{(n-1)k_-^{2}+k_+ k_-+k_+^2}}.
\end{equation}
Altogether, the total error probability per cycle is then
\begin{align}
p_\textrm{err}&=1-(1-p_\textrm{err1})(1-p_\textrm{err2})\nonumber\\
&=1-\frac{k_+^3}{(k_++N_Qk_-)(k_+^2+k_-k_++(n-1)k_-^2)}\nonumber\\
&=\frac{(1+N_Q)k_-}{k_+}+\mathcal{O}\left(\left(\frac{k_-}{k_+}\right)^2\right).
\label{perr}
\end{align}
We note, even in the high accuracy limit, that the error is sensitive to the size of the machine through the parameter $N_Q$, but note that, in a worst case scenario, is equal to $n|Q|$.

\begin{figure}

\centering\includegraphics[angle=-90,width=0.75\textwidth]{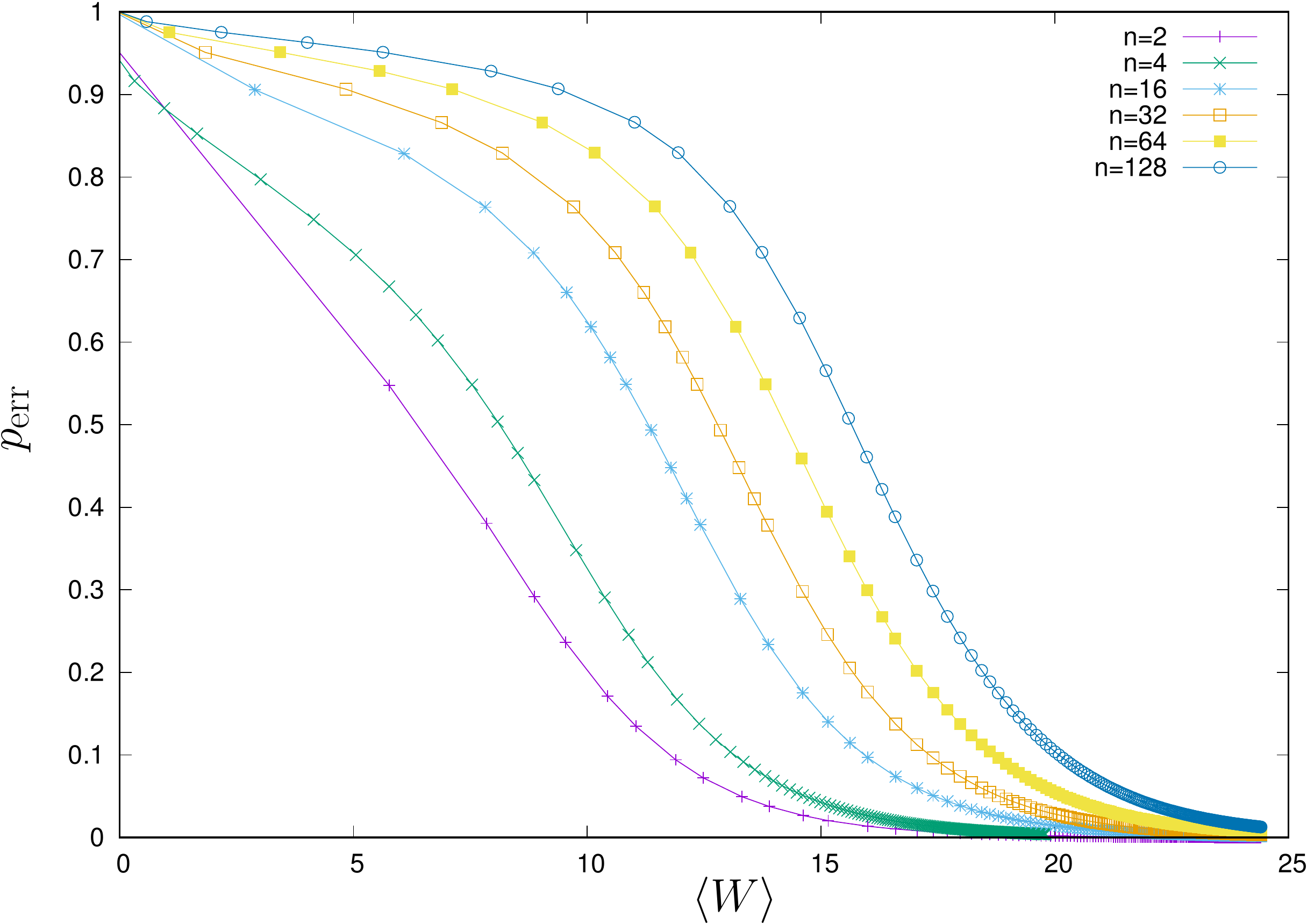}
\centering
\caption{The average work $\langle W\rangle$ to run one elementary cycle versus the error probability. For 3  different numbers of internal states $N_Q= 20,100,700$ we show the work-error relationship for alphabets of size $n=2,4,16,32,64,128$. Each dot represents a particular value of  $k_+$, which was varied from 1 to 200000; $k_-$ was kept fixed at 1.  }
\label{workerror}
\end{figure}

\subsection{Time required for the computation}

The time required for an elementary cycle to complete is spent  while waiting for the system to approach equilibrium after Step 1 and Step 3.  This time is $N_M (\tau_\textrm{set}(n)  + \tau_\textrm{flip}(N_Q))$, where $N_M$ is the chosen multiple of the relaxation time and $N_Q<n|Q|$; see equation \ref{nittime}.

\section{Discussion}

In this paper, we have presented a simple implementation of a finite state machine as a non-homogeneous continuous time Markov Chain. Key aspects of the model we have presented include the requirement for dynamic states over and above the label states that represent the internal states of the FSM that would result from a naive attempt at an implementation. Further, we have demonstrated that they can be represented using two key atomic operations to that flip and set the system amongst these states. Using this model we have investigated the energy consumption and probability of accurate computation. 
\par
We find for the FSM in the high accuracy regime that the probability of an error scales  like  $k_+^{-1}$ (see eq. \ref{perr}), while  the cost in terms of the work required to implement the cycle only increases logarithmically with $k_+$ (see eq. \ref{avgwork} and fig. \ref{workerror}). Expressed in terms of the energy differences between states $\Delta E$, we find that the average work scales (expectedly) linear with $\Delta E$ but the error falls exponentially. This benign scaling echoes previous results on different models of computation (c.f. \cite{myquasideterministic}). It tells us that  our model can compute with perfect accuracy, but only in the limit of infinite energy dissipation.   Still, quasi-deterministic computation that has practically negligible error probabilities can be done at a finite, even modest energy expense.  At the same time, it is clear that --- especially for high accuracies --- the model we proposed here dissipates well above the theoretical limit. 
\par
Interestingly, in the high accuracy limit   the size of the input  alphabet and size of the machine no longer matters for the cost of the computation; see fig. \ref{transitionfull}.  One may speculate  that a larger tape alphabet allows more information to be processed  per computational step with only marginally increased energy costs (when compared to smaller alphabets). This  suggests that it is more efficient to run Markovian computers with large alphabets than with small ones. However, the error probability can also be seen to depend, in the worst case, on the size of the machine and the size of the alphabet, $n|Q|$. The size of the alphabet can be seen to be a constant parameter in the specification of any given FSM, however the size of the machine $|Q|$ depends on the specific computation. It is important to recognise that such an error probability is only equal to $n|Q|$ in the worst case. In reality the number may be much lower than this as the value of $N_Q$ that determines the accuracy is determined by the number of incoming dynamic states in each $N$-it flip. This is an issue of the \emph{coding} of the particular algorithm to be performed by the FSM into its particular topology of internal states and transitions between them. This suggests that for any given algorithm there will be multiple possible FSM implementations, some of which will be favourable with respect to accuracy (and thus energy expenditure given some acceptable error tolerance) whilst others will be unfavourable. This opens up an interesting possibility of performance, energy and accuracy trade offs determined primarily by \emph{implementation} rather than just by the physical framework being utilised and the computation to be performed.
\par
It is important to acknowledge the limitations, simplifications and inefficiencies in the model we have presented. Firstly, when considering the efficiency of the model it is important to bear in mind that it relies on instantaneous shifts in energy levels. The energy expenditure entailed by the model  therefore does not represent a fundamental lower limit for FSMs as instantiated here. For one, the $N$-it setting could be done quasi-statically, which would reduce the cost of the running the FSM to a theoretical minimum, albeit at the cost of infinitely slow processing speed.  Having identified  $N$-it setters as the basic components required to implement FSMs, it  remains an interesting  problem to identify optimal finite time processes that can implement those, for example along the lines of \cite{deweese}.  Finally, we note the usage of infinite energy barriers in the step by step protocol that operates the FSM in this model. The use of such barriers amounts to a coarse graining of additional states and steps then inevitably ignores an energy cost and possibility of further error. In this sense we present a lower bound on the work and error calculations. However, the choice to make limited use of such energy barriers was deliberate such that dependence on fundamental aspects of FSMs, such as $|Q|$ and $n$, was retained. It is expected that the coarse graining of such processes will have progressively little effect upon the cost and error probability as the high accuracy limit $k_+\gg k_-$ is taken.

 \subsection{Programmability and entropy production}

Another property of our model is that it entails a  thermodynamic cost of programmability.   The  model proposed above  requires 4 interventions per cycle, namely lifting/lowering of state-energies and erecting barriers. The procedure can be simplified when the model of the FSM is made less general, i.e. more assumptions are made about the type of computation that will be performed.   As a first simplification  we assume that the input tape is always of a fixed length $L$. In this case, the machine can be ``unrolled,'' in the following way: Let   $\{L_{1}, L_{2}, L_{3},\ldots\}$ denote the internal states of the FSM. An unrolled FSM  whose input  tape is of length $L$, would then have  the time-indexed states $\{L_{1,1}, L_{2,1}, L_{3,1},\ldots\}$ and  $\{L_{1,2}, L_{2,2}, L_{3,2},\ldots\}$ up to  $\{L_{1,L}, L_{2,L}, L_{3,L},\ldots\}$. Additionally, it would also have for each  indexed state  corresponding   dynamic states  $\{D^{1,1}, D^{2,1}, D^{3,1},\ldots\}$.
\par
In the  unrolled FSM  transitions  always take place from a time-indexed label state to the  transition state with the same index and label. This transition is coupled to advancing the tape and a reconfiguration of the rates between transition states and dynamic states, as above, such that only those transitions are favourable that are associated with the relevant symbol from the tape alphabet.  For example, if for a given symbol the FSM stipulates a transition from, say,  state $L_i$ to state $L_j$ when the current tape element is $a$, then in the unrolled graph, this would entail a transition from $L_{i,T}$ to   $D_a^{i\to j,T}$ to $L_{j,T+1}$. All states $D_{\alpha'}^i$ with $\alpha'\neq \alpha$ would be  at a higher energy than $D_\alpha^i$.    This unrolled transition graph would have many more states than the original model, but the total number of states and the total number of transitions to be completed for one computation are known.  It is therefore possible to  arrange the  states  over $3L$ energy levels   and it is no longer necessary to erect energy barriers, which simplifies the computing protocol from 4 to 2 steps per elementary cycle. 
\par
While simpler, assuming a fixed length to the external tape limits the programmability of the system to programs that are at most of length $L$.  Based on this, we conjecture that programmability itself may entail a thermodynamic cost. This becomes even more apparent when we consider a   as a  further simplification   the case where  the precise symbol sequence of the tape, and not only its length are known. Then transition and dynamic  states are dispensable and  the machine can be reduced to state transitions between  label states only. This is because once the  input tape is known, the sequence of intended state transitions is known, and it is possible to reduce the transition graph to $L$ states arranged over $L$ energy levels.
\par
 At this point a further reduction is possible. Computationally, nothing is gained by the $L-2$ states between the initial and final state. Directly connecting the initial and final state  leads to a logically equivalent chain. Hence, once the actual input tape is known in advance, it is possible to reduce the computation to a simple bit-flip. The cost of this operation is independent of the length of the tape.

\section{Acknowledgements}

We thank Andre Barato for comments on  early versions of  this  manuscript. DC also thanks the Max Planck Institute for the Physics of Complex Systems where part of this work was done.


\end{document}